\documentclass{tlp}

\usepackage{aopmath}

\def\i#1{\hbox{\itshape #1\/}}
\def\lar{\leftarrow}
\def\nf{\i{not}\,}

\title[Strong Equivalence Made Easy]
{{Strong Equivalence Made Easy: \\
  Nested Expressions and Weight Constraints}}

\author[Hudson Turner]
{ HUDSON TURNER \\
Computer Science Department\\
University of Minnesota, Duluth \\
\email{hudson@d.umn.edu}}

\begin{document}

\bibliographystyle{acmtrans}

\maketitle

\begin{abstract}
Logic programs $P$ and $Q$ are strongly equivalent if,
given any program~$R$,
programs $P \cup R$ and $Q \cup R$ are equivalent (that is,
have the same answer sets).  Strong equivalence is convenient
for the study of equivalent transformations of logic programs:
one can prove that a local change is correct
without considering the whole program.
Lifschitz, Pearce and Valverde
showed that Heyting's logic of here-and-there
can be used to characterize strong equivalence for
logic programs with nested expressions (which subsume
the better-known extended disjunctive programs).
This note considers a simpler,
more direct characterization of strong equivalence for such
programs, and shows that it can also be applied without modification
to the weight constraint programs of Niemel{\"a} and Simons.
Thus, this characterization of strong equivalence is convenient
for the study of equivalent
transformations of logic programs written in the input languages
of answer set programming systems {\tt dlv} and {\sc smodels}.
The note concludes with a brief discussion of results that can be used to 
automate reasoning about strong equivalence, including a novel
encoding that reduces the problem of deciding the strong equivalence of a pair
of weight constraint programs
to that of deciding the inconsistency of a weight constraint program.
\end{abstract}


\section{Introduction}	\label{sec:intro}

Logic programs $P$ and $Q$ are ``strongly equivalent'' if,
given any program~$R$,
${P \cup R}$ and ${Q \cup R}$ are equivalent (that is,
have the same answer sets).
Strong equivalence is important because it allows one
to justify changes to one part of a program without considering
the whole program.  Moreover, as we show, determining that programs
are strongly equivalent is no harder (and for some classes of
programs may be easier) than determining
that they are equivalent.  

In a groundbreaking paper, Lifschitz, Pearce and Valverde (2001)
used Heyting's logic of here-and-there
to characterize strong equivalence of
logic programs with nested expressions
\cite{lif99d}.  Such ``nested programs'' subsume
the  class of extended disjunctive programs \cite{gel91b},
which can be given as input to the answer set programming system
{\tt dlv}.\footnote{{\tt Available at http://www.dbai.tuwien.ac.at/proj/dlv/ }.}

The current note characterizes strong equivalence of nested programs
in terms of concepts used in the definition
of answer sets.  Hence, no knowledge
of the logic of here-and-there is required.  In \cite{tur01a},
we showed that this characterization of strong equivalence
is easily extended to default logic \cite{rei80}.  In the
current note, we show that it
applies not only to nested programs but also to the weight constraint
programs of Niemel{\"a} and Simons (2000).  \nocite{nie00}
Thus it is also convenient for the study of equivalent
transformations of logic programs written in the input language
of the answer set programming system
{\sc smodels}.\footnote{Available at
{\tt http://www.tcs.hut.fi/Software/smodels/ }.}
We also show how to encode the question of strong equivalence
of two weight constraint programs \emph{in} a weight constraint program,
so that {\sc smodels} can be used to decide strong equivalence.

This note continues as follows.
Section~\ref{sec:lp} defines
nested programs.
Section~\ref{sec:se-lp} establishes the
characterization of strong equivalence for nested programs, and
Section~\ref{sec:lp-use} discusses strongly
equivalent transformations of them.
Section~\ref{sec:wcp} defines
weight constraint programs.
Section~\ref{sec:se-wcp} establishes the characterization
of strong equivalence for weight constraint programs, and
Section~\ref{sec:wcp-use} discusses strongly
equivalent transformations of \emph{them}.
Section~\ref{sec:relationship} compares the approach
to strong equivalence presented in this note with that
of Lifschitz, Pearce and Valverde. 
Section~\ref{sec:automation} discusses results supporting the possibility
of automated reasoning about strong equivalence, both for
nested programs (via encoding in classical propositional logic)
and for weight constraint programs (via encoding in weight constraint
programming).  \footnote{Much of the material
in Sections~\ref{sec:se-lp}, \ref{sec:lp-use} and
\ref{sec:relationship} is adapted from \cite{tur01a}.}

\section{Nested Logic Programming}  \label{sec:lp}

This paper employs the definition of nested programs
introduced in \cite{lif99d},
although the presentation differs in some details.

\subsection{Syntax}      \label{sec:lp-syntax}

The words \emph{atom} and \emph{literal} are understood here as in
propositional logic.
\emph{Elementary formulas} are literals and the 0-place connectives
$\bot$~(``false'') and $\top$~(``true''). \emph{Formulas} are built from
elementary formulas
using the unary connective $\nf$ and the binary connectives $,$
(conjunction) and $;$ (disjunction).
A \emph{rule} is an expression of the form
\begin{eqnarray*}
 && F\lar G
\end{eqnarray*}
where $F$ and $G$ are formulas, called the \emph{head} and the \emph{body} of
the rule.

A \emph{nested program} is a set of rules.

When convenient, a rule $F\lar\top$ is identified
with the formula~$F$.

A program is \emph{nondisjunctive} if the head of each rule is
an elementary formula possibly preceded by~$\nf\!$.

\subsection{Semantics}   \label{sec:lp-semantics}

Let $X$ be a consistent set of literals.

We first define recursively when~$X$ \emph{satisfies} a formula~$F$
(symbolically, $X\models F$), as follows.
\begin{itemize}
\item For elementary $F$, $X\models F$ \ iff \ $F\in X$ or $F=\top$\,.
\item $X\models(F,G)$ \ iff \ $X\models F$ and $X\models G$\,.
\item $X\models(F;G)$ \ iff \ $X\models F$ or $X\models G$\,.
\item $X\models \nf F$ \ iff \ $X\not\models F$\,.
\end{itemize}

To continue, $X$~\emph{satisfies} a rule
${F\lar G}$ if ${X\models G}$ implies ${X\models F}$,
and $X$~\emph{satisfies} a nested program~$P$ if it
satisfies every rule in~$P$.

The \emph{reduct} of a formula~$F$ relative
to~$X$ (written~$F^X$)
is obtained by replacing every
maximal occurrence in~$F$ of a formula of the form~$\nf\,G$
with $\bot$ if $X \models G$ and with $\top$
otherwise.\footnote{A maximal occurrence in~$F$ of a formula
of the form~$\nf\,G$ is: a subformula $\nf\,G$ of~$F$ such that
there is no subformula~$\nf\,H$ of~$F$ which has $\nf\,G$ as
a proper subformula.}
The \emph{reduct} of a nested program~$P$
relative to~$X$ (written~$P^X$)
is obtained by replacing the head and body of each rule in~$P$
by their reducts relative to~$X$.

Finally, $X$ is an \emph{answer set} for a nested program~$P$ if it is
minimal among the consistent sets of literals that satisfy~$P^X$.

As discussed in \cite{lif99d},
this definition agrees with previous
versions of the answer set semantics on consistent answer sets (but does
not allow for an inconsistent one).

\section{Strong Equivalence of Nested Programs}  \label{sec:se-lp}

Nested programs $P$ and $Q$ are \emph{equivalent} if they have
the same answer sets.  They are \emph{strongly equivalent} if,
for any nested program~$R$, ${P \cup R}$ and ${Q \cup R}$ are equivalent.

Notice that, as an immediate consequence, if $P$ and $Q$ are strongly equivalent,
then so are ${P \cup R}$ and ${Q \cup R}$.

\vspace*{2mm}
\noindent
\emph{Definition of SE-model} \\
For nested program~$P$,
and consistent sets~$X,Y$ of literals with ${X\subseteq Y}$,
the pair~$(X,Y)$ is an \emph{SE-model} of~$P$ if
${Y \models P}$ and ${X \models P^Y}$.\,\footnote{This definition is
stated in a slightly different form in \cite{tur01a}.  One easily verifies
that the two versions are equivalent. 
The key fact: $Y \models P$ iff $Y \models P^Y$.}

\begin{theorem}  \label{thm1} 
\cite{tur01a} 
Nested programs are strongly equivalent iff
they have the same \hbox{SE-models}.
\end{theorem}

A proof of Theorem~\ref{thm1}
is included here.
(Precisely this proof also establishes the
similar result for weight constraint programs
stated in Section~\ref{sec:wcp}.)

We begin with two lemmas.

First notice that a consistent set~$Y$ of literals
is an answer set for a program~$P$ iff $(Y,Y)$~is the unique
\hbox{SE-model} of~$P$ whose second component is~$Y$.
This observation yields the following lemma.

\begin{lemma} \label{l:1}
 Programs with the same SE-models are equivalent.
\end{lemma}

Next notice that one can decide whether a pair~$(X,Y)$ is
an \hbox{SE-model} of a program~$P$ by checking whether, for
each rule ${F \lar G}$ in~$P$, $Y$~satisfies ${F \lar G}$ and
$X$~satisfies ${F^Y \lar G^Y}$.  Hence the following.

\begin{lemma} \label{l:2}
 The SE-models of a program ${P \cup R}$ are exactly the
 SE-models common to programs~$P$ and~$R$.
\end{lemma}

The right-to-left part of the proof of Theorem~\ref{thm1} is easy
given these lemmas.  The other direction is a bit harder, but the proof
of the corresponding result in \cite{lif01}
suggests a straightforward construction which also
has the virtue of demonstrating that if nested programs~$P$ and~$Q$
are not strongly equivalent then
they can be distinguished by adding rules in which
the head is a literal and the body is either a literal or~$\top$.

\begin{proof}[Proof of Theorem~\ref{thm1}]
Right to left:  
Assume that programs~$P$ and~$Q$ have the same \hbox{SE-models}.
Take any program~$R$. We need to
show that ${P \cup R}$ and ${Q \cup R}$ are equivalent.
From Lemma~\ref{l:2} we can conclude that ${P \cup R}$ and ${Q \cup R}$ 
have the same \hbox{SE-models}, and so, by Lemma~\ref{l:1}, they
are equivalent.

Left to right:
Assume (without loss of generality)
that $(X,Y)$~is an \hbox{SE-model} of program~$P$
but not of program~$Q$.
We need to show that $P$ and $Q$ are not strongly equivalent.
Consider two cases.

\emph{Case 1}: $Y \not\models Q$.
Then ${Y \not\models Q\cup Y}$,
and so $Y$~is not an answer set for~${Q\cup Y}$.
On the other hand, since ${Y \models P}$ by assumption,
it is clear that ${Y \models P \cup Y}$.  It follows
that ${Y \models (P \cup Y)^Y}$.  Moreover, no proper
subset of~$Y$ satisfies ${(P \cup Y)^Y = P^Y \cup Y}$,
which shows that $Y$~is an answer set for~${P\cup Y}$.
Hence $P$ and $Q$ are not strongly equivalent.

\emph{Case 2}: $Y \models Q$.
Take
${R =  X \cup \{ L \leftarrow L' : L,L' \in Y \setminus X \}}$.
Clearly ${Y \models Q \cup R}$,
and it follows that ${Y \models (Q \cup R)^Y}$.
Let $Z$ be a subset of~$Y$ such that ${Z \models (Q \cup R)^Y}$
${(= {Q^Y \cup R})}$.
By choice of~$R$ we know that ${X \subseteq Z}$,
and by assumption ${X \not\models Q^Y}$,
so~${X \neq Z}$.  Hence there is some
${L \in Y \setminus X}$ that belongs to~$Z$.
It follows by choice of~$R$ that~${Y \setminus X \subseteq Z}$.
Consequently ${Z = Y}$, and so $Y$~is an answer set for~${Q \cup R}$.
On the other hand, $X$~is a proper subset of~$Y$ that
satisfies ${P^Y \cup R} = {(P \cup R)^Y}$.
So $Y$~is not an answer set for~${P \cup R}$,
and we conclude again that $P$ and $Q$
are not strongly equivalent.
\end{proof}

The form of the respective definitions may seem to suggest that 
deciding equivalence of nested programs will be easier
than deciding strong equivalence.
In fact, the opposite is (under the usual assumptions) true.
(Similar, independently-obtained complexity results appear in
\cite{pea01,lin02}.)

\begin{theorem}  \label{thm:complexity}
The problem of determining that two nested programs are equivalent
is \hbox{${\bf \Pi^P_2}$-hard}.  The problem of determining that 
they are strongly equivalent belongs to \hbox{${\bf coNP}$}.
\end{theorem}

\begin{proof}
For the first part, we show that the complementary problem
is \hbox{${\bf \Sigma_2^P}$-hard}.
Eiter and Gottlob (1993) \nocite{eit93}
showed that it is ${\bf \Sigma_2^P}$-hard
to determine that a ``disjunctive'' logic program has an
answer set.  This result extends to nested programs, which
include the disjunctive programs as a special case.
To show that a nested program is
not equivalent to the program~${\{ \bot \}}$, one must show that
it has an answer set.  So determining that two nested programs are not
equivalent is ${\bf \Sigma_2^P}$-hard.

For the second part, we observe that, given Theorem~\ref{thm1},
the complementary problem belongs to ${\bf NP}$.
That is, given two nested programs,
guess a pair $(X,Y)$ of consistent sets of literals,
and verify in polynomial time that~$(X,Y)$ is an SE-model of
exactly one of the two programs.
\end{proof}

For $\neg$-free programs (that is, programs in which $\neg$ does not occur),
the characterization of strong equivalence can
be simplified: one can restrict attention to the \emph{positive}
SE-models---those in which only atoms appear.

For any set~$X$ of literals, let $X^{+}$ be the set of atoms that belong to~$X$.

\begin{lemma} \label{l3}
Let $P$ be a $\neg$-free program.
For any consistent sets~$X,Y$ of literals
such that ${X \subseteq Y}$, $(X,Y)$~is
an SE-model of~$P$ iff $(X^{+},Y^{+})$~is.
\end{lemma}

The proof of Lemma~\ref{l3}, which is straightforward, is omitted.
The following is an easy consequence of Lemma~\ref{l3} and Theorem~\ref{thm1}.

\begin{theorem}  \label{thm4-nested}
A pair of $\neg$-free nested programs are strongly equivalent iff
they have the same positive \hbox{SE-models}.
\end{theorem}

\section{Equivalent Transformations of Nested Programs}
  \label{sec:lp-use}

To demonstrate the use of Theorem~\ref{thm1}, let us 
first consider an example discussed at length in
\cite{lif01}.  For any formulas $F$ and $G$,
programs~$P_1$ and~$P_2$ below have the same \hbox{SE-models}.
\begin{eqnarray*}
 && \begin{array}{ccc}
 F;G           & \hspace*{10mm} & F \lar \nf G \\
 \bot \lar F,G & & G \lar \nf F \\
               & & \bot \lar F,G
\end{array}
\end{eqnarray*}
To see this, take any pair~$(X,Y)$ of consistent sets of literals such that
 ${X \subseteq Y}$, and
consider four cases.
\begin{itemize}
 \item[] \emph{Case 1:} $Y \models (F,G)$.
Then ${Y \not\models P_1}$ and ${Y \not\models P_2}$,
so $(X,Y)$ is not an \hbox{SE-model} of $P_1$ or $P_2$.
 \item[] \emph{Case 2:} $Y \models (F,\nf G)$.
Then ${Y \models P_1}$ and ${Y \models P_2}$.
Since $Y \models \nf G$, ${Y \not\models G}$, and so ${Y \not\models G^Y}$.
Since $\nf$~does not occur in~$G^Y$ and ${X \subseteq Y}$,
${X \not\models G^Y}$.  We can conclude that
${X \models P_1^Y}$ iff ${X \models F^Y}$ iff 
${X \models P_2^Y}$.  So $(X,Y)$~is an SE-model of~$P_1$ iff
it is an SE-model of~$P_2$.
 \item[] \emph{Case 3:} $Y \models (\nf F,G)$.  Symmetric to previous case.
 \item[] \emph{Case 4:} $Y \models (\nf F,\nf G)$.  Similar to first case. 
\end{itemize}
It follows by Theorem~\ref{thm1} that in any program that contains~$P_1$,
$P_1$ can be safely replaced by~$P_2$, thus eliminating an occurrence
of disjunction in the heads of rules. (This result generalizes
a theorem from \cite{erd99}.)

On the other hand, Theorem~\ref{thm1} can also be used to show that no
nondisjunctive program is strongly equivalent to the
program~${\{ p;q \}}$.  (This was suggested as a challenge problem
by Vladimir Lifschitz.)
We begin with an easily verified observation.
Let $P$ be a nondisjunctive program with no occurrences of $\nf\!$.
The set of consistent sets of literals satisfying~$P$ is closed
under intersection.

\begin{proposition}
No nondisjunctive program is strongly equivalent to~$\{p;q\}$.
\end{proposition}

\begin{proof}
Let $P$ be a program strongly equivalent to~$\{p;q\}$.
Notice that both $\{p\}$ and $\{q\}$ satisfy~$\{p;q\}^{\{p,q\}}$,
but $\emptyset$ doesn't.  By Theorem~\ref{thm1}, the same is true
of~$P^{\{p,q\}}$.  It follows by the preceding observation that
$P^{\{p,q\}}$ is not nondisjunctive, and consequently neither is~$P$.
\end{proof}

Next we state a replacement theorem for nested programs.  For this
we need the following definitions.
Formulas $F$ and $G$ are \emph{equivalent}
relative to program~$P$ if, for every
SE-model~$(X,Y)$ of~$P$,
$X \models F^Y$ iff $X \models G^Y$.
An occurrence of a formula is \emph{regular} unless it is an atom
preceded by~$\neg$.

\begin{theorem}  \label{thm2}
\cite{tur01a}
Let $P$ be a nested program, and let $F$ and $G$ be
formulas equivalent relative to~$P$. 
For any nested program~$Q$, and any
nested program~$Q'$ obtained from~$Q$
by replacing regular occurrences of~$F$ by~$G$,
programs~${P \cup Q}$ and ${P \cup Q'}$ are strongly equivalent.
\end{theorem}

The restriction to regular occurrences is essential.  For example,
formulas~$p$ and~$q$ are equivalent relative to program
${P_3 = \left\{ p \lar q,\; q \lar p \right\}}$,
yet programs ${P_3 \cup \{ \neg p \}}$ and
${ P_3 \cup \{ \neg q \} }$ are not strongly equivalent.

Theorem~\ref{thm2} is a more widely-applicable version of Proposition~3 from
\cite{lif99d}.
There we defined equivalence of formulas more strictly, and did not
make it relative to a program.  We also used a notion of ``equivalence''
of programs stronger than strong equivalence.
Many formula equivalences are proved there
(see Proposition~4, \cite{lif99d}),
and of course they also hold
under this new (weaker) definition (relative to the empty program).
Thus, Theorem~\ref{thm2} implies, for instance, that replacing subformulas
of the form~${\nf (F,G)}$ with ${\nf F; \nf G}$ yields a strongly
equivalent program.

For another example using Theorem~\ref{thm2},
observe that for any program~$Q$, and any program~$Q'$ obtained from~$Q$
by replacing occurrences of~$\nf F$ by~$G$ and/or~$\nf G$ by~$F$,
programs~$P_2 \cup Q$ and $P_2 \cup Q'$ are strongly equivalent.

\section{Weight Constraint Programming}  \label{sec:wcp}

This presentation is adapted from \cite{fer01}, and
extends slightly the definition of weight constraint programming
from \cite{nie00}.

\subsection{Syntax}

A \emph{rule element} is a literal (\emph{positive} rule element)
or a literal prefixed with $\nf$ (\emph{negative} rule element).
A \emph{weight assignment} is an expression of the form
\begin{eqnarray}
 && e = w
\label{wa}
\end{eqnarray}
where $e$ is a rule element and $w$ is a nonnegative real number (a ``weight'').
The part ``$=w$'' of~(\ref{wa}) can be omitted if $w$ is~$1$.
A \emph{weight constraint} is an expression of the form
\begin{eqnarray}
 && L \leq S \leq U
\label{wc}
\end{eqnarray}
where $S$~is a finite set of weight assignments, and
each of~${L,U}$ is a real number or one of the symbols
$-\infty$, $+\infty$.
The part ``$L \leq$'' can be omitted from~(\ref{wc}) if $L$ is~$-\infty$;
similarly, the part ``$\leq U$'' can be omitted if $U$ is~$+\infty$.
A \emph{WCP rule} is an expression of the form
\begin{eqnarray}
 && C_0 \lar C_1,\ldots,C_n
\label{wcp-rule}
\end{eqnarray}
where $C_0,\ldots,C_n$ ($n \geq 0$) are weight constraints such that
$C_0$~does not contain negative rule elements.  We call $C_0$ the \emph{head}
of~(\ref{wcp-rule}), and
the rule elements that
occur in the head are called the \emph{head literals} of~(\ref{wcp-rule}).

A \emph{weight constraint program} is a set of WCP rules.

This syntax becomes a generalization of the syntax of ``extended''
logic programs, introduced in \cite{gel90},
if we allow a rule element~$e$ to stand for the
weight constraint~${1 \leq \{e\}}$.

In weight constraint programs, let~$\bot$ stand for
the weight constraint~${1 \leq \{\}}$.
When convenient, a WCP rule ${C \lar}$ is identified with the
weight constraint~$C$.

\subsection{Semantics}

Let $X$ be a consistent set of literals.

For any finite set~$S$ of weight assignments, let
\begin{eqnarray*}
 && v(S,X) =
  \sum_{\scriptstyle e= w \,\in\, S \atop \scriptstyle X \models e}\!\!\! w\,.
\end{eqnarray*}
We say $X$ \emph{satisfies} a weight constraint $L \leq S \leq U$
if $L \leq v(S,X) \leq U$.
To continue, $X$ satisfies a WCP rule ${C_0 \lar C_1,\ldots,C_n}$
if $X$ satisfies~$C_0$ whenever $X$ satisfies all of ${C_1,\ldots,C_n}$,
and $X$ satisfies a weight constraint program~$P$
if it satisfies every rule in~$P$.

For any weight constraint that can be written in the form ${L \leq S}$,
its \emph{reduct} $(L \leq S)^X$ with respect to~$X$ is the
weight constraint~${L^X \leq S'}$ where
\begin{itemize}
 \item $S' = \{ e=w \in S\ |\ \hbox{$e$ is a positive rule element}\}$, and
 \item $L^X = L - v(S\setminus S',X)$\,.
\end{itemize}
The \emph{reduct} of a WCP rule
\begin{eqnarray}
 && L_0 \leq S_0 \leq U_0 \lar 
        L_1 \leq S_1 \leq U_1,\ldots,L_n \leq S_n \leq U_n
\label{WCP-rule-long}
\end{eqnarray}
with respect to~$X$ is the weight constraint program
consisting of all rules
\begin{eqnarray*}
 && e \lar (L_1 \leq S_1)^X,\ldots,(L_n \leq S_n)^X
\end{eqnarray*}
such that
\begin{itemize}
 \item $e$~is a head literal of~(\ref{WCP-rule-long}),
 \item ${X \models e}$, and
 \item $X \models S_i \leq U_i$ for all ${i \in \{1,\ldots,n\}}$.
\end{itemize}
The \emph{reduct} $P^X$ of a weight constraint program~$P$
with respect to~$X$ is
the union of the reducts with respect to~$X$ of all rules in~$P$.

Finally, $X$ is an answer set for a weight constraint program~$P$
if $X$~satisfies~$P$ and no proper subset of~$X$
satisfies~$P^X$.\,\footnote{The 
requirement that $X \models P$, which does not appear explicitly in the
definition of answer sets for nested programs,
is necessary here because, for weight constraint programs,
$X \models P^X$ does not imply that $X \models P$.  Note that
the converse \emph{does} still hold.}

The semantics of nested programs and weight constraint programs agree
wherever their syntax overlaps.

\section{Strong Equivalence for Weight Constraint Programs}
                                              \label{sec:se-wcp}

The definitions of equivalence, strong equivalence and
SE-models for weight constraint programs are
as they were for nested programs.  

\begin{theorem} \label{thm3}
Weight constraint programs are strongly equivalent iff
they have the same \hbox{SE-models}.
\end{theorem}

As mentioned previously, the proof of Theorem~\ref{thm1} applies,
without change, to this theorem also.\footnote{This easy correspondence implies that
there is no difficulty in allowing a more general class of
programs that can be formed as the union of a nested program and
a weight constraint program.  No doubt more elaborate hybrids are
possible as well, some of them quite straightforward.}

In addition, as with nested programs,
the proof of Theorem~\ref{thm3} shows that
weight constraint programs that are not
strongly equivalent can be distinguished by adding rules that either
can be represented by a literal
or can be written ${L \lar L'}$, where $L$ and~$L'$ are literals.

Moreover, it is clear that the {\bf coNP} complexity
of deciding strong equivalence carries over to weight constraint programs.
(The same easy argument applies.)

And also as with nested programs, when programs are $\neg$-free (that is,
have no occurrences of~$\neg$), we can
restrict attention to positive SE-models (in which only atoms appear).
Lemma~\ref{l3}, stated previously in the context of nested programs,
holds also for weight constraint programs, and together with
Theorem~\ref{thm3} yields:

\begin{theorem}  \label{thm4}
A pair of $\neg$-free
weight constraint programs are strongly equivalent iff
they have the same positive \hbox{SE-models}.
\end{theorem}

\section{Equivalent Transformations of Weight Constraint Programs}  
                                                 \label{sec:wcp-use}

We start with an adaptation of the first example from Section~\ref{sec:lp-use}.
For any literals~$L$ and~$L'$,
the two programs shown below have the same \hbox{SE-models}.
\begin{eqnarray*}
 && \begin{array}{ccc}
  L \lar \nf L' & \hspace*{10mm} & 1\leq\{L,L'\}\leq 1\\
  L' \lar \nf L \\
  \bot \lar L,L'
\end{array}
\end{eqnarray*}
This is easily verified, much as was done for the similar nested program example.

Ferraris and Lifschitz (2001) introduced a translation from
weight constraint programs to nested programs, and argued that this
translation is interesting in part because it provides, indirectly,
a method for reasoning about strong equivalence of weight constraint
programs.  Next we consider the main example from that paper.

We are interested in the n-Queens program consisting of the following rules,
where ${i,i',j,j' \in \{1,\ldots,n\}}$.
\begin{eqnarray}
 && 1 \leq \{q(1,j),\ldots,q(n,j)\} \leq 1 
\label{q1} \\
%for each column, a queen in exactly one row
 && 1 \leq \{q(i,1),\ldots,q(i,n)\} \leq 1
\label{q2} \\
%for each row, a queen in exactly one column
 && 
 \bot \lar q(i,j),q(i',j') \qquad (|i\!-\!i'|=|j\!-\!j'|)
\label{q3}
\end{eqnarray}
Intuitively, (\ref{q1}) expresses that, for each column~$j$, 
there is a queen in exactly one
row.  Similarly, (\ref{q2}) expresses that, for each row~$i$, 
there is a queen in exactly one
column.  Finally, (\ref{q3}) stipulates that no two queens occupy a common diagonal.

We wish to verify that (\ref{q2})~can be equivalently replaced by
the following.
\begin{eqnarray}
 && 
 \bot \lar q(i,j),q(i,j') \qquad (j < j')
\label{q2'}
\end{eqnarray}
What we'll show is that program~$P$ consisting of rules~(\ref{q1}) and (\ref{q2}) is
strongly equivalent to program~$Q$ consisting of rules~(\ref{q1}) and (\ref{q2'}).

Observe first that the positive SE-models of the rules~(\ref{q1})
are exactly the pairs~$(X,X)$ where
\begin{eqnarray*}
 & & X = \{q(i_1,1),\ldots,q(i_n,n)\}
\end{eqnarray*}
with $i_1,\ldots,i_n \in \{1,\ldots,n\}$.
To see this, notice that 
\begin{itemize}
 \item the rules~(\ref{q1}) are satisfied by all and only such sets~$X$, and
 \item the reduct of these rules with respect to such an~$X$ is not satisfied by
   any proper subset of~$X$ (and in fact can be written as~$X$).
\end{itemize}
Next observe that such an $X$ satisfies the
rules~(\ref{q2}) iff all of $i_1,\ldots,i_n$ are different.
Finally, such an $X$ satisfies the rules~(\ref{q2'}) under exactly the
same conditions.  We can conclude that
programs $P$ and $Q$ have the same positive \hbox{SE-models}, and
by Theorem~\ref{thm4} they are strongly equivalent.

\section{SE-Models and the Logic of Here-and-There}
                                        \label{sec:relationship}

Lifschitz, Pearce and Valverde (2001)
identify $\neg$-free nested logic program rules with
formulas in Heyting's logic of here-and-there,
and show that $\neg$-free nested programs are strongly equivalent
iff they are equivalent in the logic of here-and-there.
They also explain that this result can be extended to all nested programs
(including those in which $\neg$ occurs)
in the standard fashion, by translating a program with occurrences
of $\neg$ into one without. (See their paper for details.)

According to their definitions,
an {\sl HT-interpretation} is a pair~$(I^H,I^T)$ of sets
of atoms, with ${I^H \subseteq I^T}$.  Without going into details, we can
observe that they define when an HT-interpretation is a model of
a $\neg$-free nested program in the sense of the logic of here-and-there.
Although it is not done here, one can
verify that their Lemmas~1 and~2
together imply the following.

\begin{proposition} \label{prop1}
For any $\neg$-free nested program~$P$,
$(X,Y)$~is a positive SE-model of~$P$ 
iff $(X,Y)$~is a model of~$P$ in the logic
of here-and-there.
\end{proposition}

Not surprisingly, it then follows from Theorem~\ref{thm4-nested} that
these two characterizations of strong equivalence are essentially equivalent
with regard to $\neg$-free nested logic programs.  A key advantage though of
the SE-models approach is that it is easily extended to other, similar
nonmonotonic formalisms.  In Section~\ref{sec:se-wcp}, the SE-models
characterization of strong equivalence
was extended to weight constraint programs (without altering
the definition of SE-model or the proof of the strong equivalence
theorem).
In \cite{tur01a},
a similar notion of SE-model was used to characterize strong
equivalence for default logic.
The extension in this case was
also easy.\footnote{In fact, although we do not go into
details here, the SE-models characterization of strong equivalence
is also easily adapted to the causal theories formalism of \cite{mcc97}
and the modal causal logic UCL of \cite{tur99}.  Similar characterizations are
likely for other nonmononotic formalisms based on such fixpoint semantics.}

Even when we consider strong
equivalence only for nested programs, it seems that
the SE-models and
the here-and-there characterizations have different strengths.

One advantage of the SE-models approach is its
relative simplicity.  The definition is quite straightforward,
based on concepts already introduced in the definition of answer
sets.  This in turn simplifies the proof of the strong equivalence
theorem.  

The definition of an SE-model for nested programs
takes advantage of the
special status of the symbol~$\lar$ in usual definitions of logic programming.
By comparison, the logic of here-and-there treats $\lar$ as
just another connective, and even defines $\nf$ in terms of it---${\nf F}$
is understood as an abbreviation for ${\bot \lar F}$.
The possibility of nested occurrences of~$\lar$ complicates
the truth definition considerably.

It is important to note, though, that
this complication takes a familiar form---the 
truth definition in the logic of here-and-there uses standard Kripke models.
In fact, they are a special case of Kripke models for intuitionistic logic
(which is, accordingly, slightly weaker).  Thus, such an
approach brings with it a range of associations that
may help clarify intuitions about the meaning of connectives $\lar$
and $\nf$ in logic programming.

Even if we consider only convenience in the study
of strong equivalence (or similar properties),
the logic of here-and-there
offers a potential advantage: it is a logic with
known identities, deduction rules, and such, which can be
used to reason about strong equivalence in particular cases.

Nonetheless, when we wish to apply strong equivalence results,
it seems likely that a model-theoretic
argument based on SE-models will often be easier than
a proof-theoretic argument using known properties of the logic
of here-and-there.

\section{Toward automated reasoning about strong equivalence}
                                            \label{sec:automation}

It may be desirable to use
automated methods to reason about strong equivalence.
One possibility would be to employ
general-purpose tools for the logic of here-and-there, but
we may instead wish to take advantage of recent results regarding encodings of 
strong equivalence of nested programs in classical propositional
logic \cite{pea01,lin02}.  The existence of such encodings is suggested
by the {\bf coNP} complexity of the decision problem, and indeed they
are not hard to find.\footnote{More surprising
is a related result due to Lin (2002) showing that
strong equivalence of disjunctive logic programs with variables and constants
(but without proper functions) is also {\bf coNP}, despite the fact
that equivalence for such programs is undecidable!}
An encoding like those in \cite{pea01,lin02} is specified below.
After this, we specify a similar, new encoding of strong equivalence of
weight constraint programs, this time \emph{in} a weight constraint program.

\subsection{Strong equivalence of nested programs as unsatisfiability} %(in classical propositional logic)}

The key is a translation
that maps a nested program to a propositional theory whose models
are in one-to-one correspondence with the SE-models of the
program.\footnote{The easy proof in Lin's paper (for disjunctive
programs only) is based directly on this idea.
Pearce et al.\  (2001) show that the same kind of
translation can be applied to any theory in the logic of here-and-there,
yielding a classical propositional theory whose models correspond
to the models of the original theory (in the logic of here-and-there).}
Here it is convenient to restrict consideration to $\neg$-free programs.

Consider any $\neg$-free nested program~$P$.
The first step is to augment the signature---for each
atom~$A$ of the language of~$P$ 
add a new atom~$A'$.  A classical interpretation~$I$
of the augmented language corresponds to a pair~$(I_{new},I_{old})$,
where $I_{old}$ consists of the original atoms that are true in~$I$
and $I_{new}$ consists
of the original atoms~$A$ such that ${I \models A'}$.
Then for any rule ${F \lar G}$, let ${pl(F \lar G)}$ stand for 
${pl(G) \supset pl(F)}$, where $pl(F)$ and $pl(G)$ are obtained from~$F$ and
$G$ by replacing occurrences of~$\nf$ with~$\neg$, $;$~with~$\vee$ and
$,$~with~$\wedge$, yielding a formula of classical logic.
For any classical propositional formula~$\phi$, let $\phi'$~be obtained
from~$\phi$ by replacing each occurrence of an 
atom~$A$ of the original language that is
not in the scope of~$\neg$ with its counterpart~$A'$.
Let $pl(P)$ be the following classical propositional theory.
$$\{ pl(r) : r \in P \} \cup \{pl(r)' : r \in P\} \cup
\{ A' \supset A: \hbox{$A$ is an atom of the original language}\}$$
It is straightforward to verify that ${I \models pl(P)}$ iff 
$(I_{new},I_{old})$~is an SE-model of~$P$.  Moreover, every
positive SE-model of~$P$ can be written in the form~$(I_{new},I_{old})$
for some interpretation~$I$ of the language of~$pl(P)$.

Given this encoding of the positive SE-models of a $\neg$-free
nested program, we can construct (via Theorem~\ref{thm4-nested}),
for any finite $\neg$-free nested programs~$P$ and~$Q$,
a classical propositional formula that is satisfiable iff 
$P$~and~$Q$ are not strongly equivalent.  Abusing notation, take $\i{pl}(P)$
to stand for the (finite) conjunction of its elements,
and do the same for~$\i{pl}(Q)$.
Then $P$ and $Q$ are strongly equivalent iff
the formula $\i{pl}(P)\not\equiv\i{pl}(Q)$ is unsatisfiable.

\subsection{Strong equivalence of weight constraint programs
as inconsistency (in weight constraint programming)}

One could devise a similar encoding in classical propositional logic
for the \hbox{SE-models} of a weight constraint program (and then decide
strong equivalence of a pair of finite weight constraint programs
by deciding unsatisfiability
of a classical propositional formula, as above).  Unfortunately
this would require a translation of weight constraints
into classical propositional logic, which would in general be rather costly.
In light of this, it may be preferable instead to use an encoding \emph{in}
a weight constraint program.  The crucial step---capturing the
SE-models of a weight constraint program as answer sets of a weight
constraint program---is quite easy.  But the subsequent step---encoding
the equivalence of two weight constraint programs as inconsistency
of a weight constraint program---requires some additional work,
as we will see.
We again restrict consideration to $\neg$-free programs.  

Consider any $\neg$-free weight constraint program~$P$. 
Augment the language as before, using similar
notation~$(X_{new},X_{old})$ for the pair
of sets of atoms in the original language
corresponding to a set~$X$ of atoms in the augmented language.
(So ${X = X_{old} \cup \{ A' : A \in X_{new}\}}$.)
Let $P'$~be obtained from~$P$ by
replacing each occurrence of an atom~$A$ not preceded by~$\nf$ with~$A'$.
Let $wc(P)$ be the program obtained by adding to ${P \cup P'}$ the rules
\begin{eqnarray*}
 && \bot \lar A',\nf A\,, \\
 && \{ A\, \}\,, \\
 && \{ A'\, \}\,,
\end{eqnarray*}
for each atom~$A$ of the original language.  Notice that $X$~is an answer set
for~$wc(P)$ iff ${X \models wc(P)}$, since inclusion of the
rules of forms $\{ A\, \}$ and $\{ A'\, \}$ effectively renders every
atom in the language of $wc(P)$ abducible.   Notice also that, as before, 
it is
straightforward to verify that ${X \models wc(P)}$ iff 
$(X_{new},X_{old})$ is an SE-model of~$P$, and that, moreover, every
positive SE-model of~$P$ can be written~$(X_{new},X_{old})$
for some subset~$X$ of the atoms of the language of~$wc(P)$.

Given this encoding of the positive SE-models of a $\neg$-free
weight constraint program, Theorem~\ref{thm4} allows us
to reduce the problem of deciding the
strong equivalence of $\neg$-free weight constraint programs~$P$ and~$Q$
to the problem of deciding the equivalence of weight constraint programs
$\i{wc}(P)$ and $\i{wc}(Q)$.  We conclude by showing
that this latter question can, in turn, 
be encoded in weight constraint programming.

Recently, Janhunen and Oikarinen \nocite{jan02} (2002)
investigated such encodings, but their results
do not cover all programs we are interested in. 
On the other hand, our programs
$\i{wc}(P)$, $\i{wc}(Q)$ are unusual:
their answer sets are simply the sets of atoms that satisfy them,
because they make all atoms abducible (so to speak).
So for our purposes it will be sufficient to describe an
encoding (in weight constraint programming) of the following question:
Are two $\neg$-free weight constraint programs (in the
same language) satisfied by exactly the same sets of atoms?

To this end, we will first define a transformation that takes any 
$\neg$-free weight constraint program~$P$ to a program~$\i{not}(P)$
whose answer sets, roughly speaking,
correspond to the sets of atoms that do not satisfy~$P$.
Let ${\bf A}$ denote the set of all atoms in the language of~$P$.
The language of~$\i{not}(P)$ is obtained by adding to~${\bf A}$
a new atom~$\i{witness}$, as well as a new
atom~$h(C_0)$ for each weight constraint~$C_0$ that appears at least
once as the head of a rule in~$P$.  
For each rule ${C_0 \lar C_1,\ldots,C_n}$ of~$P$,
program~$\i{not}(P)$ includes the rules
\begin{eqnarray}
 && h(C_0) \lar C_0\,, \label{r1} \\
 && \i{witness} \lar \nf h(C_0),C_1,\ldots,C_n\,.  \label{r2}
\end{eqnarray}
Program $\i{not}(P)$ also includes, for every atom~$A \in {\bf A}$, the rule
\begin{eqnarray}
 && \{A\,\}       \label{r3}
\end{eqnarray}
and finally the rule
\begin{eqnarray}
 && \bot \lar \nf \i{witness} \,. \label{r4}
\end{eqnarray}
Notice that the rules~(\ref{r3}) make the atoms in~${\bf A}$ abducible,
while the rules~(\ref{r1}) and~(\ref{r2}) can provide support only for
atoms not in~${\bf A}$.
In light of these observations it is not difficult to verify that,
for every subset~$X$ of~${\bf A}$,
the program consisting just of rules~(\ref{r1}),
(\ref{r2}) and (\ref{r3}) will have an answer set 
obtained from~$X$ by adding: (i)~the atom~$h(C_0)$ for each head~$C_0$
from~$P$ such that ${X \models C_0}$, and (ii)~the atom \i{witness} if
there is a rule from~$P$ that $X$ does not satisfy.
Moreover, all
answer sets of (\ref{r1})--(\ref{r3}) can be obtained in this way.
The effect of adding rule~(\ref{r4}) then is just to eliminate those
answer sets for which~${X \models P}$.
Consequently, for every subset~$X$ of ${\bf A}$,
${X \not\models P}$ iff there is an answer set~$Y$ for~$not(P)$
such that ${Y \cap {\bf A} = X}$.

Now consider a second $\neg$-free weight constraint
program~$Q$ in the same language
as~$P$.  Again because program~$\i{not}(P)$
makes all atoms in ${\bf A}$ abducible,
we can conclude that, for any subset~$X$ of~${\bf A}$,
${X \not\models P}$ and ${X \models Q}$ 
iff there is an answer set~$Y$ for~$\i{not}(P) \cup Q$
such that ${Y \cap {\bf A} = X}$.
It follows that $P$ and~$Q$ are satisfied by exactly the same sets of atoms
iff both ${not(P) \cup Q}$ and ${P \cup not(Q)}$ are inconsistent 
(that is, have no answer sets).

As a last step in this construction,
it is straightforward to combine two weight constraint programs into
a single program that is consistent iff at least one of the original two is.
Again let's call these programs~$P$ and~$Q$.  Add new atoms $p$ and $q$ to
their common language.  Add $p$ to the body of each rule in~$P$;
add $q$ to the body of each rule in~$Q$.  Take the resulting
rules and add to them
one more rule: ${1 \leq \{p,q\} \leq 1}$.  Let's call the resulting program
$\i{or}(P,Q)$.  It has an answer set iff at least one of $P$ and $Q$ does.

So, summarizing the result of this subsection,
we can decide strong equivalence of arbitrary
$\neg$-free weight constraint programs~$P$ and~$Q$ (in the same language)
by deciding the inconsistency of the $\neg$-free weight constraint program
\begin{eqnarray*}
\i{or}\left(\i{not}(\i{se}(P)) \cup \i{se}(Q), 
          \i{se}(P) \cup \i{not}(\i{se}(Q))\right)\,.
\end{eqnarray*}
That is, this program has no answer sets iff $P$ and $Q$ are strongly equivalent.

\section*{Acknowledgements}

Many thanks to Vladimir Lifschitz, who showed me drafts of his
related papers and encouraged me to write up what I noticed.
Thanks to Alessandro Provetti for a useful conversation about
automated reasoning about strong equivalence,
and to the anonymous referees for helpful suggestions.
This work partially supported by NSF Career Grant \#0091773.


\begin{thebibliography}{}

\bibitem[\protect\citeauthoryear{Eiter and Gottlob}{Eiter and
  Gottlob}{1993}]{eit93}
{\sc Eiter, T.} {\sc and} {\sc Gottlob, G.} 1993.
\newblock Complexity results for disjunctive logic programming and application
  to nonmonotonic logics.
\newblock In {\em Logic Programming: Proceedings of the 1993 International
  Symposium}. 266--278.

\bibitem[\protect\citeauthoryear{Erdem and Lifschitz}{Erdem and
  Lifschitz}{1999}]{erd99}
{\sc Erdem, E.} {\sc and} {\sc Lifschitz, V.} 1999.
\newblock Transformations of logic programs related to causality and planning.
\newblock In {\em Logic Programming and Non-monotonic Reasoning: Proc.~Fifth
  Int'l Conf. (Lecture Notes in Artificial Intelligence 1730)}. 107--116.

\bibitem[\protect\citeauthoryear{Ferraris and Lifschitz}{Ferraris and
  Lifschitz}{2001}]{fer01}
{\sc Ferraris, P.} {\sc and} {\sc Lifschitz, V.} 2001.
\newblock Weight constraints as nested expressions.
\newblock Available at {\tt www.cs.utexas.edu/users/vl/papers.html}.

\bibitem[\protect\citeauthoryear{Gelfond and Lifschitz}{Gelfond and
  Lifschitz}{1990}]{gel90}
{\sc Gelfond, M.} {\sc and} {\sc Lifschitz, V.} 1990.
\newblock Logic programs with classical negation.
\newblock In {\em Logic Programming: Proc. of the 7th Int'l Conference},
  {D.~Warren} {and} {P.~Szeredi}, Eds. 579--597.

\bibitem[\protect\citeauthoryear{Gelfond and Lifschitz}{Gelfond and
  Lifschitz}{1991}]{gel91b}
{\sc Gelfond, M.} {\sc and} {\sc Lifschitz, V.} 1991.
\newblock Classical negation in logic programs and disjunctive databases.
\newblock {\em New Generation Computing\/}~{\em 9}, 365--385.

\bibitem[\protect\citeauthoryear{Janhunen and Oikarinen}{Janhunen and
  Oikarinen}{2002}]{jan02}
{\sc Janhunen, T.} {\sc and} {\sc Oikarinen, E.} 2002.
\newblock Testing the equivalence of logic programs under stable model
  semantics.
\newblock In {\em Logics in Artificial Intelligence: Proc.\ 8th European
  Conference (JELIA'02)}. 493--504.

\bibitem[\protect\citeauthoryear{Lifschitz, Pearce, and Valverde}{Lifschitz
  et~al\mbox{.}}{2001}]{lif01}
{\sc Lifschitz, V.}, {\sc Pearce, D.}, {\sc and} {\sc Valverde, A.} 2001.
\newblock Strongly equivalent logic programs.
\newblock {\em ACM Transactions on Computational Logic\/}~{\em 2}, 526--541.

\bibitem[\protect\citeauthoryear{Lifschitz, Tang, and Turner}{Lifschitz
  et~al\mbox{.}}{1999}]{lif99d}
{\sc Lifschitz, V.}, {\sc Tang, L.}, {\sc and} {\sc Turner, H.} 1999.
\newblock Nested expressions in logic programs.
\newblock {\em Annals of Mathematics and Artificial Intelligence\/}~{\em
  25,\/}~2--3, 369--390.

\bibitem[\protect\citeauthoryear{Lin}{Lin}{2002}]{lin02}
{\sc Lin, F.} 2002.
\newblock Reducing strong equivalence of logic programs to entailment in
  classical logic.
\newblock In {\em Proc.~of KR'02}. 170--176.

\bibitem[\protect\citeauthoryear{McCain and Turner}{McCain and
  Turner}{1997}]{mcc97}
{\sc McCain, N.} {\sc and} {\sc Turner, H.} 1997.
\newblock Causal theories of action and change.
\newblock In {\em Proc.~of AAAI-97}. 460--465.

\bibitem[\protect\citeauthoryear{Niemel{\"a} and Simons}{Niemel{\"a} and
  Simons}{2000}]{nie00}
{\sc Niemel{\"a}, I.} {\sc and} {\sc Simons, P.} 2000.
\newblock Extending the {\sc smodels} system with cardinality and weight
  constraints.
\newblock In {\em Logic-Based Artificial Intelligence}, {J.~Minker}, Ed.
  Kluwer, 491--521.

\bibitem[\protect\citeauthoryear{Pearce, Tompits, and Woltran}{Pearce
  et~al\mbox{.}}{2001}]{pea01}
{\sc Pearce, D.}, {\sc Tompits, H.}, {\sc and} {\sc Woltran, S.} 2001.
\newblock Encodings for equilibrium logic and logic programs with nested
  expressions.
\newblock In {\em Proc.~of the 10th Portuguese Conf.~on AI (Lecture Notes in
  Artificial Intelligence 2258)}. 306--320.

\bibitem[\protect\citeauthoryear{Reiter}{Reiter}{1980}]{rei80}
{\sc Reiter, R.} 1980.
\newblock A logic for default reasoning.
\newblock {\em Artificial Intelligence\/}~{\em 13,\/}~1,2, 81--132.

\bibitem[\protect\citeauthoryear{Turner}{Turner}{1999}]{tur99}
{\sc Turner, H.} 1999.
\newblock A logic of universal causation.
\newblock {\em Artificial Intelligence\/}~{\em 113}, 87--123.

\bibitem[\protect\citeauthoryear{Turner}{Turner}{2001}]{tur01a}
{\sc Turner, H.} 2001.
\newblock Strong equivalence for logic programs and default theories (made
  easy).
\newblock In {\em Logic Programming and Nonmonotonic Reasoning: Proc.\ of Sixth
  Int'l Conf.\ (Lecture Notes in Artificial Intelligence 2173)}. 81--92.

\end{thebibliography}
\end{document}